\documentclass[fleqn,usenatbib]{mnras}

\usepackage{newtxtext,newtxmath}

\usepackage[T1]{fontenc}

\DeclareRobustCommand{\VAN}[3]{#2}
\let\VANthebibliography\thebibliography
\def\thebibliography{\DeclareRobustCommand{\VAN}[3]{##3}\VANthebibliography}

\usepackage{graphicx}	
\usepackage{amsmath}	
\usepackage{lscape}
\usepackage{lineno}
\usepackage{threeparttable}
\usepackage{url}
\usepackage{hyperref}

\title[Ionized Wind in Mrk 6]{Ionized X-ray winds in the radio galaxy Mrk 6}

\author[T. Kayanoki et al.]{
Taishu Kayanoki,$^{1}$\thanks{E-mail: kayanoki@astro.hiroshima-u.ac.jp}
Junjie Mao,$^{2,1}$
Yasushi Fukazawa$^{1}$
\\
$^{1}$Department of Physics, Graduate School of Advanced Science and Engineering, Hiroshima University 1-3-1 Kagamiyama, Higashi-Hiroshima,\\ Hiroshima 739-8526, Japan\\
$^{2}$Department of Physics,  Tsinghua University, Beijing 100084, China\\
}

\date{Accepted XXX. Received YYY; in original form ZZZ}

\pubyear{2023}

\begin{document}
\label{firstpage}
\pagerange{\pageref{firstpage}--\pageref{lastpage}}
\maketitle

\begin{abstract}
Active galactic nucleus (AGN) outflows including jets and ionized winds have been key phenomena such as jet collimation and AGN feedback to the host galaxy in astrophysics. 
Radio galaxies, a type of AGN with misaligned jets, have provided valuable insights into the properties and relationships of these outflows.
However, several aspects regarding AGN outflows remain unclarified, such as the relationship between jets and ultrafast outflows (UFOs) and the differences between the properties of radio-loud AGN disk winds and radio-quiet AGN ionized winds.
To clarify these aspects, radio galaxies containing UFOs and warm absorbers (WAs) must be studied.
Currently, both UFOs and WAs have been reported in only two radio galaxies, 3C 390.3 and 4C $+$74.26.
To enhance our understanding of the connection between the jets and ionized winds, we conducted a study on Mrk 6, a potential candidate for the third case of a radio galaxy displaying these characteristics.
Using X-ray spectra obtained from {\it XMM-Newton}, we performed photoionization modeling using the SPEX code.
The best-fit model analysis results revealed the presence of a UFO component with a relatively low ionization parameter (Fe {\sc xix}/{\sc xviii} lines blueshifted by $-34700^{+400}_{-200}~{\rm km~s^{-1}}$) and a WA component with an outflow velocity of $-7600 \pm 200~{\rm km~s }^{-1}$.
To further confirm the nature of the UFO and WA in Mrk 6, we simulated the X-ray imagining and spectroscopy mission spectra.
\end{abstract}




\section{Introduction}
Active galactic nuclei (AGNs) accrete mass onto supermassive black holes (SMBHs) while rejecting outflow as jets or ionized winds.
Ionized winds are characterized by their ionization parameter $\xi$, hydrogen column density $N_{\rm H}$, and outflow velocity, and are classified as ultrafast outflows \citep[UFOs,][]{Tombesi2010}, warm absorbers \citep[WAs,][]{Crenshaw2003, Halpern1984} and obscures \citep[e.g.][]{Kaastra2014, Mehdipour2017, Mao2022}.
Ionization $\xi$ is defined as $\xi = L_{\rm ion} / n_{\rm H}~r^2$, where $L_{\rm ion}$ ($1-1000$ Ryd) denotes the ionizing luminosity, $n_{\rm H}$ represents the disk wind hydrogen number density, and $r$ indicates the distance from the source \citep{Tarter1969}.
Generally, UFOs are observed mainly as blueshifted absorption lines at $7-9$ keV \citep{Tombesi2010}.
The UFO ionization parameter is $\log [\xi ({\rm erg~cm~s^{-1}})]=3-5$, the hydrogen column density is $\log [N_{\rm H} ({\rm cm^{-2}})]=22-23.5$, the outflow velocity is $v_{\rm out} = 10,000-70,000~{\rm km~s^{-1}}$ \citep{Laha2021}, and the kinetic power of the mass outflow is $\sim 0.1-10\%$ of Eddington luminosity \citep[e.g.][]{Gofford2015,Tombesi2012}.
As such, UFOs are estimated to exist in $\sim40\%$ AGNs \citep{Tombesi2010, Tombesi2011, Gofford2013}.
Based on simulations \citep[e.g.][]{Nomura2016},UFOs are launched at $\sim 30$ times the Schwarzschild radius $R_{\rm S} = 2GM_{\rm BH} / c^2$, and from observations \citep[e.g.][]{Gofford2015}, UFOs originate at $10^2-10^4~R_{\rm S}$.
Although prior research suggested that UFOs may originate from the jet base \citep{Ghisellini2004}, the launch radius of UFOs remains unclear and the relationship between the jet and UFO has not yet been characterized.
In general, WAs are detected mainly as blueshifted absorption lines or edges in the soft X-ray spectrum.
The WAs ionization parameter is $\log [\xi ({\rm erg~cm~s^{-1}})]=-1-3$, the hydrogen column density is $\log [N_{\rm H} ({\rm cm^{-2}})]=21-22.5$, the outflow rate is $v_{\rm out} = 100-2,000~{\rm km~s^{-1}}$ \citep{Sako2001,Laha2016}.
WAs exist at $0.1-1000$ pc, and their hydrogen number density $n_{\rm H}$ has been estimated to be $n_{\rm H} \sim 10^{4}-10^{11}~{\rm cm}^{-3}$ \citep[][and references therein]{Laha2021}.
Recently, the possibility of a new classification of entrained ultra-fast outflows \citep[E-UFO, e.g.][and references therein]{Serafinelli2019}, which have fast outflow velocity, low-ionization, and low-density, has been suggested.

AGNs can be classified by radio intensity (or radio loudness).
The radio intensity $R$ is defined based on \citet{Kellermann89} as $R=F_{\rm 6~cm}/F_{\rm opt}$ using the radio flux ($F_{\rm 6~cm}$) and the optical flux $\log F_{\rm opt} = (48.36-B)/2.5$ \citep{Schmidt83}.
For radio-loud AGN, $R\geq10$, and for radio-quiet AGN, $R<10$.
Although the spectral energy distribution (SED) of radio-loud and radio-quiet AGNs are similar across the infrared to X-ray spectrum, their radio and gamma radiations differ owing to the strong relativistic jets emitted by radio-loud AGNs \citep[e.g.][]{Urry1995, Kayanoki22}.
In principle, radio galaxies are radio-loud AGNs with jets misaligned to the line of sight, which facilitate the observation of the jets as well as the central core in such galaxies.
Therefore, radio galaxies are deemed as the most suitable objects for investigating the environment surrounding SMBHs and characterizing the relationship between the jet and accretion disk and that between the jet and disk winds.
Recently, several scholars have conducted statistical studies on radio galaxies, e.g., X-ray absorption in radio galaxies in \citet{Kayanoki22}, UFOs in radio galaxies in \citet{Tombesi2010a}, \citet{Tombesi2012b}, \citet{Gofford2013}, \citet{Tombesi2013} and \citet{Tombesi2014}, and WAs in radio galaxies in \citet{Reeves2009}, \citet{Torresi2010}, \citet{Torresi2012}, \citet{Tombesi2016}, and \citet{Mehdipour19}.
Notably, WAs are seldom observed in radio galaxies and might be related to the power of the jets \citep{Mehdipour19}; however, a direct physical relationship has not yet been established.

Jets are presumed to be magnetically powered; however, the collimation of jets in the vicinity of SMBHs is not understood.
A quasi-analytical study suggested that disk winds may promote the collimation of jets near SMBHs \citep{Fukumura2014, Globus2016, Blandford2019}.
Moreover, as reported by \citet{Mehdipour19}, WAs are related to the jet intensity (radio-loudness).
However, several questions regarding the AGN outflows remain unsolved, such as the possible variations in the properties of disk winds in radio-loud AGNs and radio-quiet AGNs, including the relationship between jets and UFOs.
To resolve these questions, the AGN with observed jets, UFOs, and WAs, i.e., radio galaxies containing both UFOs and WAs, provide vital insights.
However, only two radio galaxies have been reported to contain both UFOs and WAs, 3C 390.3 and 4C $+$74.26 (\cite{Tombesi2014} for UFOs, \cite{Mehdipour19} for WAs).
Therefore, understanding the physical state of the disk winds in radio galaxies is crucial for resolving these problems.
Mrk 6 is the Seyfert 1.5-type AGN in a lenticular galaxy \citep{Kharb14}.
$B$-band magnitude and 6 cm radio flux ($F_{\rm 6~cm}$) of Mrk 6 is $B=15.16$ and $F_{\rm 6~cm} = 0.115$ Jy \citep{Veron2010}, Mrk 6 can be classified as a radio-loud galaxy ($R = 30$).
Specifically, it has a black hole mass of $(1.8\pm0.2)\times 10^8 M_\odot$ \citep{Doroshenko2012} with $z = 0.019$ \citep{Vaucouleurs1991}.
In \citet{Mehdipour19}, the X-ray data analysis of the European Photon Imaging Camera \citep[EPIC,][]{Struder2001, Turner2001} and Reflection Grating Spectrometer \citep[RGS,][]{Herder2001} detectors on the {\it XMM-Newton} satellite for 16 radio galaxies reported WAs, among which Mrk 6 exhibited the highest density of WAs. 
Furthermore, \citet{Schurch2006} reported that Mrk 6 contains disk winds or an ionized skin or an atmosphere of the molecular torus.
However, as the UFO of the Mrk 6 has not yet been discussed, this study analyzed the Mrk6 disk winds in more detail. 
The cosmological parameters ($H_0$,~$\Omega_m$,~$\Omega_\Lambda$) used in this study are $H_0=70~\rm{km~s^{-1}Mpc^{-1}}$, $\Omega_M=0.3$, $\Omega_{\Lambda}=0.7$.

\section{Observation and data reduction}
Mrk 6 was observed with {\it XMM-Newton} in March 2001, April 2003, and October 2005 (Table \ref{mrk6}).
In this study, we used the RGS data for the soft X-ray spectra and the positive-negative junction (pn) charged-coupled device (CCD) camera \citep{Struder2001} of the EPIC for the hard X-ray spectra, because EPIC/MOS data does not significantly increase the statistics.
The {\it XMM-Newton} data were reprocessed and analyzed using Science Analysis System version 19.1.0 and the calibration data from current calibration files (CCF) were updated on November 18, 2021.
After eliminating the data at instances unsuitable for analysis because of background flares, the exposure durations of the EPIC detector were shorter than those of the RGS detector data (Table \ref{mrk6}).
Moreover, the observations (IDs 0061540101 and 0305600501) were unsuitable for disk wind analysis, because their extremely short exposure durations produce excessively large errors in the spectral data points. 
Herein, the analysis was performed on the data with the longest EPIC/pn exposure time, i.e., observation ID 0144230101. 

\begin{table}
 \caption{XMM-Newton observation of Mrk 6. }
 \label{mrk6}
 \centering
  \begin{tabular}{ccccc}
\noalign{\smallskip}\hline\noalign{\smallskip}
Observation ID & \multicolumn{3}{c}{Net Exposure (ks)} & Date\\
 & RGS1 & RGS2 & EPIC/pn & \\
\noalign{\smallskip}\hline\noalign{\smallskip}
0061540101 & 33.8 & 24.9 & 10.5 & 2001-03-27\\
0144230101 & 34.4 & 34.4 & 31.5 & 2003-04-26\\
0305600501 & 21.0 & 21.0 & 4.49 & 2005-10-27\\
\noalign{\smallskip}\hline\noalign{\smallskip}
\end{tabular}
\end{table}

For the EPIC/pn data reduction, the {\tt epchain} pipeline task was used.
In EPIC/pn ($>0.50~{\rm counts~s^{-1}}$), periods with high flaring were filtered out by applying the {\tt \#XMMEA\_EP} filters.
The EPIC/pn spectra were extracted from a circular region of the center coordinate (06:52:11.6, +74:25:35.1) with a radius of 35.750 arcsec. 
Moreover, the background was extracted from a neighboring source-free rectangle region of center coordinates (06:52:05.0, +74:22:05.9), length and width of 96 arcsec and 262 arcsec, respectively, and a tilt of 342.29$^{\circ}$ in the same CCD. 
Overall, the pile-up was evaluated to be negligible.
Thereafter, the response matrices were generated for the EPIC/pn spectra of each observation using the {\tt rmfgen} and {\tt arfgen} tasks. 
For the RGS data, the {\tt rgsproc} pipeline task was used for data reduction, and the source and background spectra were extracted to generate the response matrices.
Moreover, the time intervals with background count rates $>3\sigma$ were filtered out with the Chandra Interactive Analysis of Observations \citep[CIAO,][]{Fruscione2006} version 4.14 {\tt lc\_sigma\_clip} task.
To rectify the cross-calibration between the EPIC/pn and RGS spectra, the RGS1 and RGS2 spectra were rescaled by a factor of 0.91 and 0.89, respectively, to correspond to the flux level in the $1.2-1.3$ keV energy range.

\section{Spectral Fitting and Result}
In this study, the spectral analysis was performed using the SPEX code v3.06.01 \citep{Kaastra2020}.
For spectral analysis, EPIC/pn spectra were used in the energy range of $1.77-10.0$ keV, and the {\tt obin} command of the SPEX code was applied to perform optimal binning according to \citet{Kaastra2016}.
The two RGS spectra were utilized for the $7.0-27.0$ \AA~wavelength range and binned by a factor of 2 to ensure that the bin size was approximately 3rd of the RGS energy resolution.
In particular, the spectral fitting was performed on the RGS and EPIC/pn spectra using the {\it C}-statistic \citep{Kaastra2017}.
In all the fittings, performed hereinafter, all models were multiplied by the neutral absorption model ({\tt hot} model) to account for the absorption by our Galaxy. 
The column density of our Galaxy was set to $N_{\rm H} = 7.63\times10^{20}~{\rm cm}^{-2}$ \citep{HI4PI}.
For the reflection and absorption models,  we considered the protosolar abundances of \citet{Lodders2009}.
The error for each parameter refers to the 68.3\% confidence level.

\subsection{Search for emission and absorption lines}
\begin{figure}
 \centering
 \includegraphics[width=0.9\hsize]{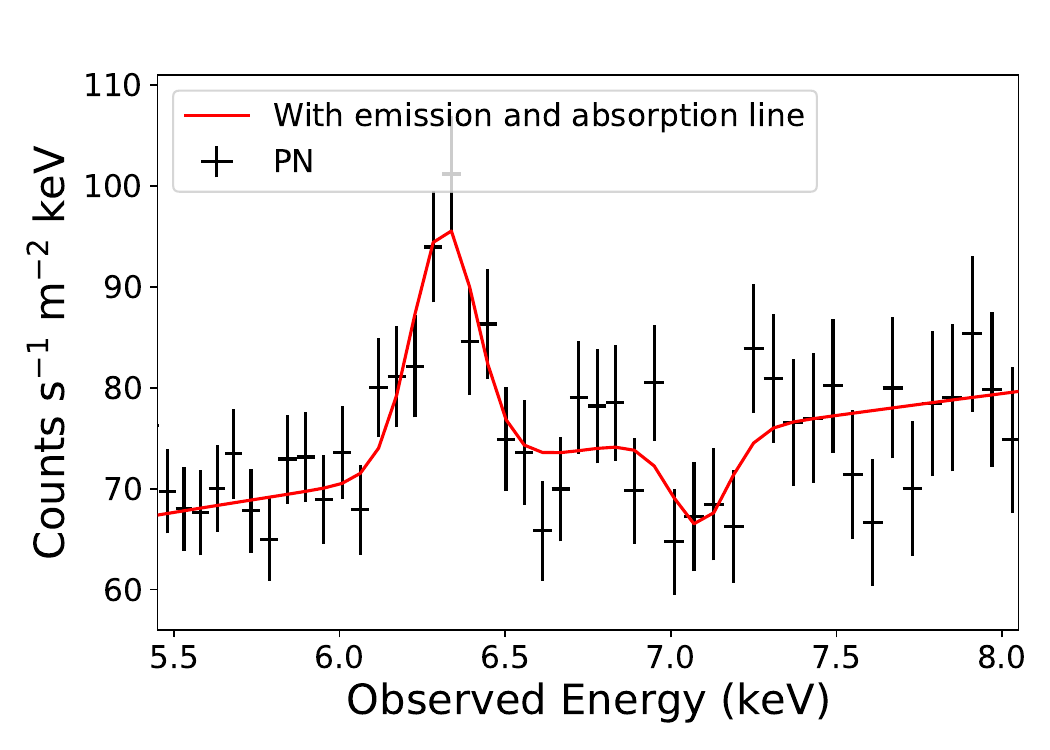}

 \caption{EPIC/pn spectra in $5-10$ keV energy band and the power-law + emission line + absorption line model; black cross point and red line represent EPIC/pn spectra and power-law + gaussian (emission line) + gaussian (absorption line), respectively. For the EPIC/pn spectra we used the {\tt obin} command for optimal binning. }
 \label{gabs}
\end{figure}

First, to inspect the emission and absorption lines in the spectra, we analyzed the EPIC/pn spectra in the $5-10$ keV energy band using the power-law ({\tt pow}) model, the power-law ({\tt pow}) + emission line model, and the power-law ({\tt pow}) + emission line ({\tt gaus}) + absorption line ({\tt gaus}) model (Table \ref{abs_check}).
The EPIC/pn spectra in the $5-10$ keV energy band and the power-law ({\tt pow}) + emission line ({\tt gaus}) + absorption line model ({\tt gaus}) are illustrated in Figure \ref{gabs}.
The Fe emission line at 6.4 keV and absorption line at $\sim 7$ keV were observed in Figure \ref{gabs}.
In addition to Table \ref{abs_check}, the model with the absorption lines displayed a more accurate {\it C}-statistic ($\Delta C=-5.9$, this would be a marginal detection) in Figure \ref{gabs}, thereby suggesting the existence of UFOs.
The emission line at 6.4 keV was considered to be fluorescence X-rays owing to the neutral or low ionized iron in AGN, whereas the $\sim 7$ keV absorption lines were considered to be caused by a highly ionized gas.

\begin{table}
 \caption{Emission line and absorption line search}
 \label{abs_check}
 \centering
 \hspace{-1cm}
 \begin{threeparttable}
 \begin{tabular}{lccc}
\noalign{\smallskip}\hline\noalign{\smallskip}
& power-law & power-law & power-law \\
&  & + emission line & + emission line \\
&  &  & + absorption line \\
\noalign{\smallskip}\hline\noalign{\smallskip}
{\tt pow} $\Gamma$ $^{(1)}$ & $1.61\pm0.05$ & $1.59\pm0.05$ & $1.58^{+0.05}_{-0.02}$\\
{\tt pow} Norm $^{(2)}$ & $2.9^{+0.3}_{-0.2}$ & $2.7^{+0.3}_{-0.2}$ & $2.7^{+0.1}_{-0.2}$\\
{\tt gaus} e $^{(3)}$ & - & $6.43\pm0.02$ & $6.43\pm0.02$\\
{\tt gaus} norm $^{(4)}$ & - & $13\pm2$ & $12\pm2$\\
{\tt gaus} FWHM $^{(5)}$ & - & $130\pm10$ & $120\pm10$\\
{\tt gaus} e $^{(3)}$ & - & - & $7.22\pm0.02$\\
{\tt gaus} norm $^{(4)}$ & - & - & $-3\pm1$\\
{\tt gaus} FWHM $^{(5)}$ & - & - & $<160$\\
\noalign{\smallskip}\hline\noalign{\smallskip}
Total {\it C}-stat / d.o.f. & 149.9 / 83 & 82.4 / 80 & 76.5 / 77\\
Total Expected {\it C}-stat & 85.1$\pm$13.1 & 85.1$\pm$13.1 & 85.1$\pm$13.1\\
$\Delta${\it C}-stat & - & -67.5 &  -73.4\\
\noalign{\smallskip}\hline\noalign{\smallskip}
\end{tabular}
\begin{tablenotes}
\item
Fit the EPIC/pn hard X-ray data (5-10 keV).
$\Delta${\it C}-stat: Difference in C-statistics between the simple power-law model.
{\tt gauss} model used for obtaining emission and absorption lines.
(1) Photon index of the X-ray power-law component.
(2) Normalization of the X-ray power-law component in $10^{51}~{\rm photons~s^{-1}~keV^{-1}}$ at 1 eV.
(3) Line energy in keV.
(4) Normalization of Gaussian in $10^{48} ~{\rm photons~s^{-1}}$.
(5) Line FWHM in keV.
\end{tablenotes}
\end{threeparttable}
\end{table}

\subsection{Continuum}\label{Continuumsection}
Furthermore, we investigated the features (photon index and flux) of the continuum component of the X-ray spectra.
As the high-energy band is considered to be weakly affected by absorption, the EPIC/pn data were fitted with a model in which the power law was multiplied by the absorption ({\tt hot}) by the neutral gas of the host galaxy.
In addition, as the $6.4$ keV Fe emission lines possibly existed based on the results of the previous subsection, the power-law ({\tt pow}) + reflection component ({\tt refl}) multiplied by the absorption (hot) of the neutral gas of host galaxy was fitted to the EPIC/pn data.
Moreover, the RGS and EPIC/pn data were fitted with a model that multiplied the modified blackbody component ({\tt mbb}) + power-law ({\tt pow}) + reflection component ({\tt refl}) by the absorption ({\tt hot}) of the neutral gas in the host galaxy.

Compared to the power-law model with the power-law ({\tt pow}) + reflection component ({\tt refl}) model, the {\it C}-statistic for the total spectrum (EPIC/pn + RGS spectrum) is more suitable for the power-law model without the reflection component ($\Delta$ {\it C} = 100.5). 
Nonetheless, the {\it C}-statistic for the EPIC/pn data spectrum is more appropriate for the model with the reflection component. 
A comparison of the {\it C}-statistics of the EPIC/pn data spectra indicated the considerably superior performance of the model incorporating the reflective component.
This is because the reflection model can adequately explain the $6.4$ keV Fe emission lines.
However, the power-law ({\tt pow}) + reflection component ({\tt refl}) model does not adequately represent the RGS (soft X-ray spectrum).
Accordingly, the modified blackbody component ({\tt mbb}) was added to the continuum.
As listed in Table \ref{comt_tau2}, the {\it C}-statistic for the total spectrum (EPIC/pn + RGS spectrum) was 2286.8, which is a relatively more accurate result.
Based on the aforementioned findings, we adopted the modified blackbody model ({\tt mbb}) + power-law ({\tt pow}) + reflection ({\tt refl}) as the continuum component.
In the following subsection, we investigated the suitability of the neutral gas {\tt hot} model \citep{Plaa2004, Steenbrugge2005} and the ionized gas {\tt pion} model \citep{Miller2015, Mehdipour2016} as the absorbed component.

\subsection{Absorption component}
We investigated whether the neutral gas ({\tt hot}) model or the ionized gas ({\tt pion}) model (disk wind model) is appropriate as the absorber.
As discussed in Section \ref{Continuumsection}, the M1 model multiplying the neutral gas ({\tt hot}) model to the continuum did not explain the $\sim7$ keV absorption line.
Therefore, we adopted the photoionization model {\tt pion}.
The photoionization equilibrium of the {\tt pion} model is calculated self-consistently using the available plasma routines of SPEX \footnote{ \url{https://spex-xray.github.io/spex-help/models/pion.html\#sec-pion} }.
Therefore, the {\tt pion} model could calculate the SED and SED filtering (the second PION sees a filtered SED) with the photoionization equilibrium simultaneously.
However, the photoionization equilibrium of the other photoionization model (e.g. the {\tt xabs} model) is pre-calculated for a grid of $\xi$-values by an external code, for instance, Cloudy or XSTAR.
For the {\tt pion} model, the hydrogen density were set to 1 ${\rm cm}^{-3}$ and the broadening velocity were set to 100 ${\rm km~s^{-1}}$.
These two values are the default values of the {\tt pion} model.

\begin{figure}
 \centering
 \includegraphics[width=0.9\hsize]{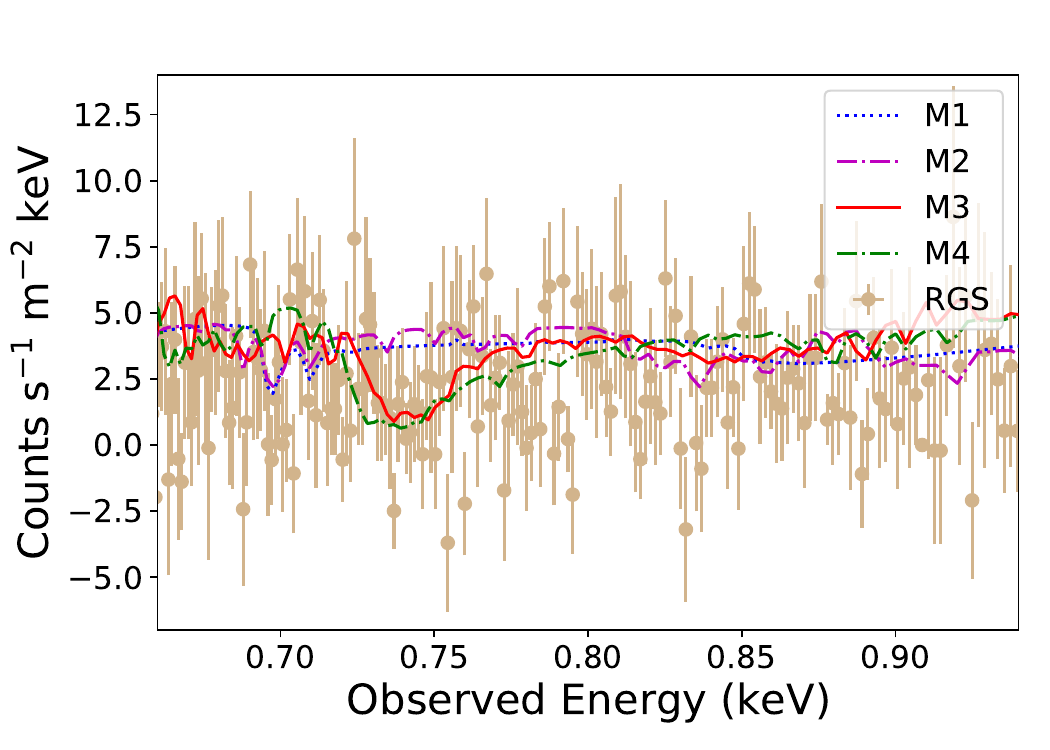}
 \caption{Soft X-ray spectra (RGS) and model curve, wherein the orange point, blue dotted line, magenta dash-dotted line, red line, and green line represent the combined RGS spectra, M1 model, M2 model, M3 model, and M4 model, respectively. RGS spectra were combined by {\tt rgscombine} command and the binning factor was set to 5.}
 \label{abssoft}
\end{figure}

\begin{figure}
 \centering
 \includegraphics[width=0.9\hsize]{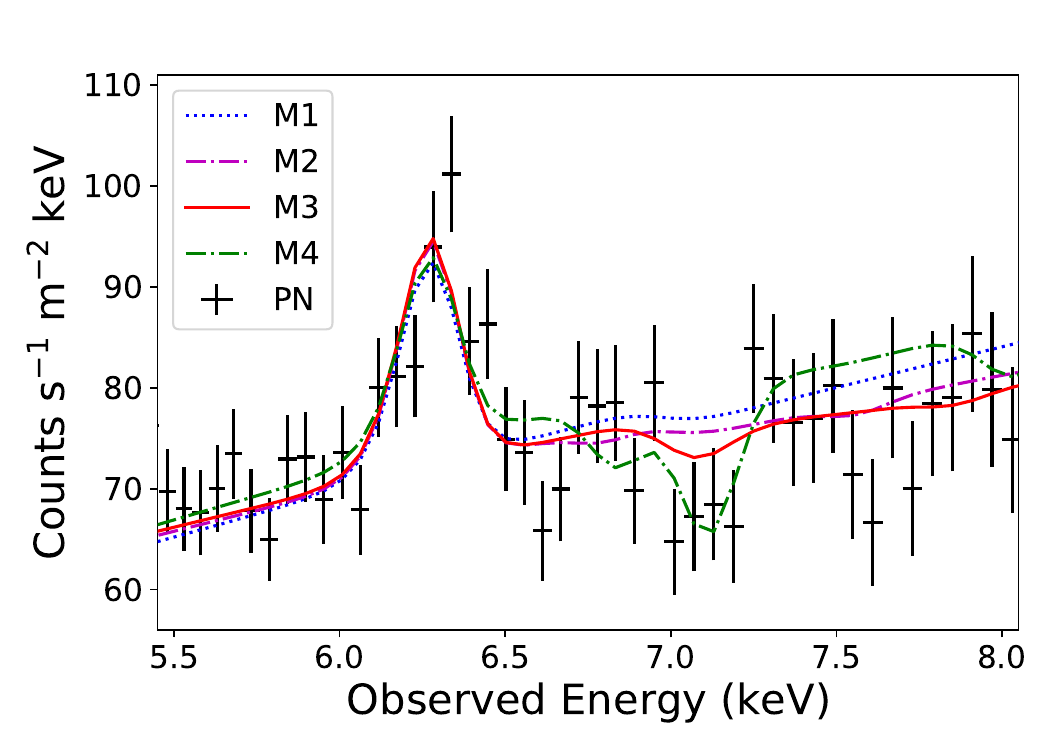}
 \caption{Energy band spectra (EPIC/pn)of $5-10$ keV and model curve, wherein the black cross point, blue-dotted line, magenta dash-dotted line, red line, and green line represent the EPIC/pn spectra, M1 model, M2 model, M3 model, and M4 model, respectively. For the EPIC/pn spectra, the {\tt obin} command was used for optimal binning.}
 \label{absfe}
\end{figure}

First, we added a single {\tt pion} component to the M1 model (M2 model).
Compared to the M1 model, the M2 model improved the {\it C}-statistic.
However, the ionizing absorption component of the M2 model exhibited soft X-ray absorption and did not represent the $\sim~7$ keV absorption line.
Subsequently, the neutral gas ({\tt hot}) component in the M2 model was modified to an ionized gas ({\tt pion}) component (M3 model).
Initially, the M3 model was assumed to contain one neutral gas ({\tt hot}) and two ionized gases. 
As the column density for the {\tt hot} model was obtained as zero from the fitting results, the M3 model potentially did not require a neutral gas ({\tt hot}).
Compared to the M2 model, the M3 model improved the {\it C}-statistic by 61.7.
The M3 model exhibited the soft X-ray absorption (Figure \ref{abssoft}) and $\sim~7$ keV absorption lines (however, the $\sim~7$ keV absorption lines were weak (Figure \ref{absfe}).
Although the $>$7 keV absorption lines of UFOs are typically considered to be Fe {\sc xxv}/{\sc xxvi} absorption lines \citep[e.g.][]{Tombesi2014}. 
However, the M3 model has a low ionization parameter and it ascribed the $\sim7$ keV absorption lines to Fe {\sc xix}/{\sc xviii}.
Therefore, the M3 model has weak absorption lines.
Thus, the parameters of the M3 model were modified and fitted again.
The obtained result may be a local minimum (M4 model).
Although the M4 model more accurately represented the $\sim7$ keV absorption lines (Figure \ref{absfe}), its C-statistic of 36.6 was inferior to that obtained with the M3 model.
However, the $\sim7$ keV absorption lines can be possibly attributed to the Fe {\sc xxv}/{\sc xxvi} absorption lines, the spectra obtained herein suggested that the $\sim7$ keV absorption lines are Fe {\sc xix}/{\sc xviii}.
As such, the data from the X-ray imagining and spectroscopy mission (XRISM) can be utilized to verify whether the $\sim7$ keV absorption lines in Mrk 6 are caused by Fe {\sc xxv}/{\sc xxvi} or Fe {\sc xix}/{\sc xviii}.
The fitting results for the M1, M2, M3, and M4 models are summarized in Table \ref{comt_tau2}.

At this point, we tried thawing the covering factor and Fe abundance of the UFO component for both the M3 model and the M4 model.
First, we thawed the covering factor (${\rm C_f}$) of the UFO component for the M3 model and the M4 model.
For the M3 model, the covering factor changed to ${\rm C_f}$ = 0.7 and the C-stat improved significantly ($\Delta C = -4.3$).
On the other hand, for the M4 model, the covering factor did not change significantly.
Therefore, we thawed the covering factor model for the M3 model.
Next, we thawed the Fe abundance of the UFO component for the M3 model and the M4 model.
For the M3 model, the Fe abundance did not change significantly.
On the other hand, for the M4 model, the Fe abundance changed to 3.3 solar and the C-stat improved significantly ($\Delta C = -4.7$).
As a result, the M3 model has the lowest C-stat value (Table \ref{comt_tau2}) among the above 4 cases.
Therefore, the M3 model with a covering factor of 0.7 was deemed to be the best model.

\begin{figure*}
  \begin{minipage}[b]{0.45\hsize}
    \centering
    \includegraphics[scale=0.4]{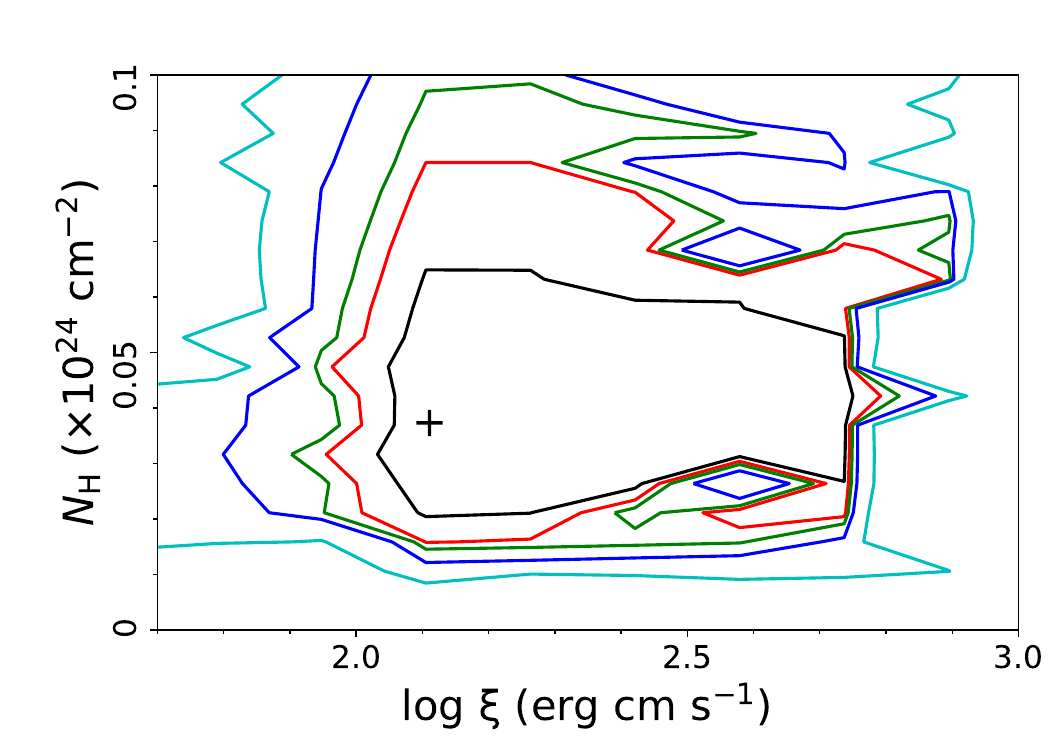}
  \end{minipage}
\begin{minipage}[b]{0.45\hsize}
    \centering
    \includegraphics[scale=0.4]{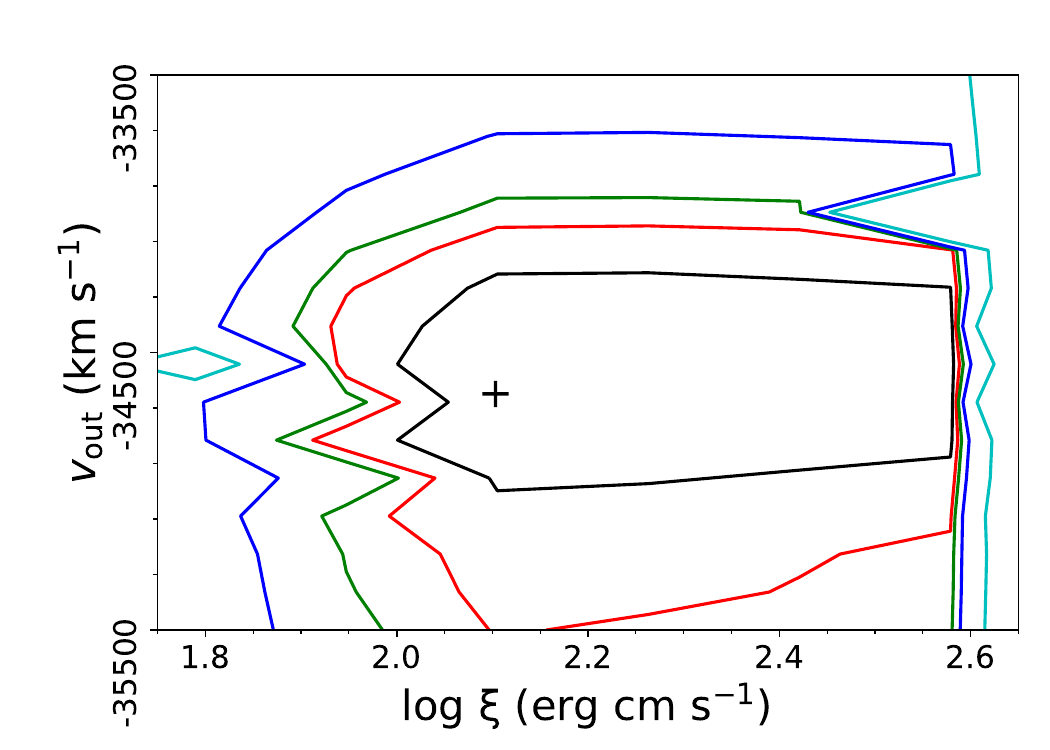}
  \end{minipage}\\
  \begin{minipage}[b]{0.45\hsize}
    \centering
    \includegraphics[scale=0.4]{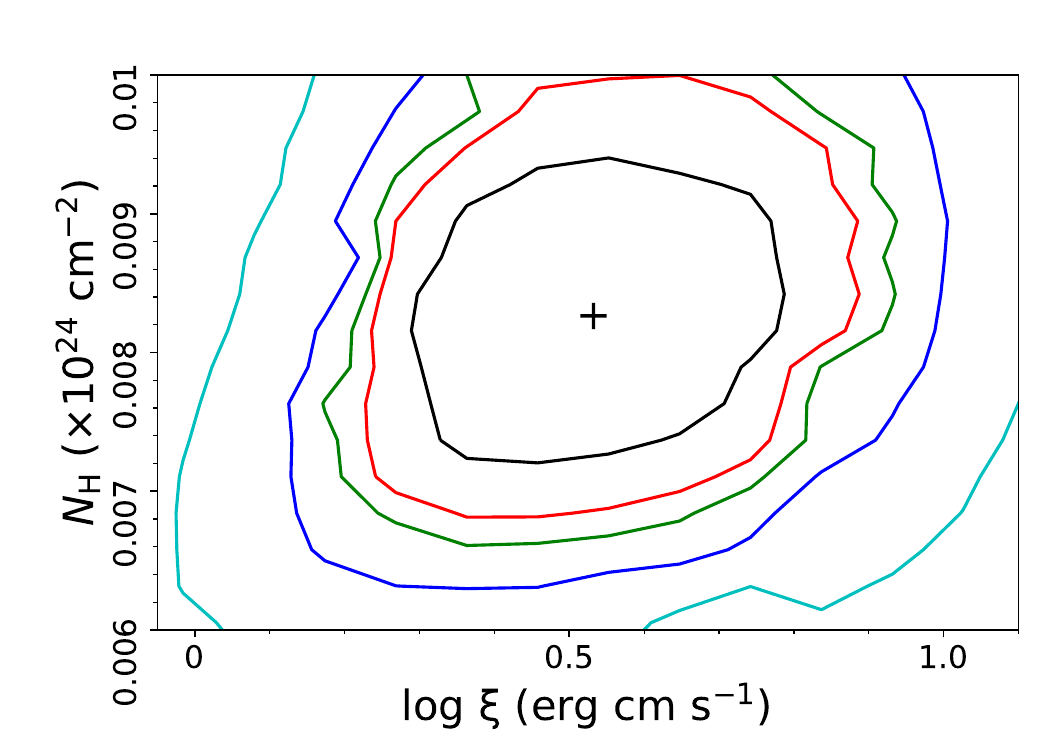}
  \end{minipage}
\begin{minipage}[b]{0.45\hsize}
    \centering
    \includegraphics[scale=0.4]{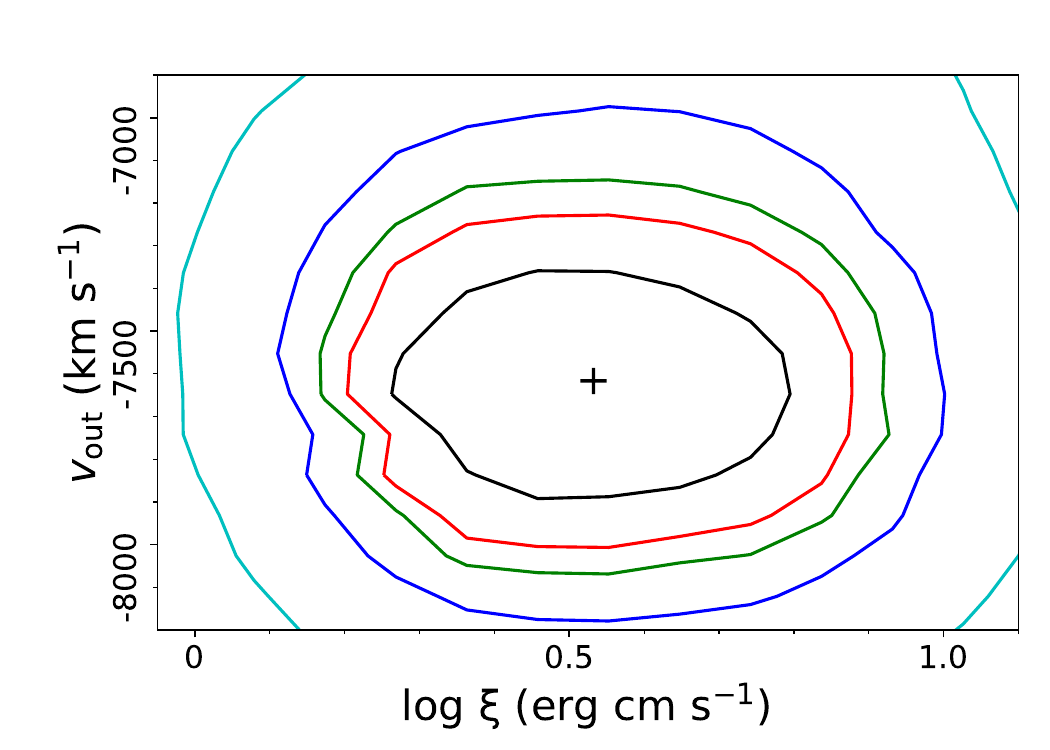}
  \end{minipage}
  \caption{Contour maps based on the M3 model (upper left panel: the ionization parameter vs. column density of UFO component. upper right panel: ionization parameter vs. outflow velocity of UFO component. bottom left panel: the ionization parameter vs. column density of WA component. bottom right panel: ionization parameter vs. outflow velocity of WA component.). Black cross point, black line, red line, green line, blue line, and cyan line show the best-fit parameter, 68.3\% confidence level, 90\% confidence level, 95.4\% confidence level, 99\% confidence level, and 99.99\% confidence level, respectively.}\label{p2acontour}
\end{figure*}

To investigate whether the M3 model and M4 model are not a local minimum, we varied the value of outflow velocity in the range $0-0.35~c$ while performing the fitting and checked the C-stat.
In this test, both the M3 model and the M4 model look not local minimum. 
(We could not determine whether the M4 model is just a local minimum or not.)
In addition, we made contour maps for the M3 model of the UFO component (Fig \ref{p2acontour}).
We did the likelihood ratio test to estimate the statistical confidence level of the M3 model with PN spectra.
The likelihood ratio is (PN C-stat of the M2 model) $-$ (PN C-stat of the M3 model) = 183.0 $-$ 166.8 = 16.2.
PN spectra have 162 bins, and then we decided that we could use the likelihood ratio test for PN spectra and approximate the test following the $\chi^2$ distribution.  
Since the difference in degrees of freedom between the M2 model and the M3 model is 3, the $p$-value of the M3 model should be a $p<0.001$.
From the above, the confidence level of the M3 model is $>99.9\%$.

In addition, to investigate the number of layers of the ionized gas, the M3 model was multiplied by an additional component of the ionized gas ({\tt pion}) model (M5 model).
Similar to the M3 model, the column density of the neutral absorption component was zero in this model.
Thus, the neutral absorption ({\tt hot}) component was not required in the M5 model.
Compared to the M3 model, the M5 model improved the {\it C}-statistic by only 7.9 with the 3 degrees of freedom.
Thus, the third ionized gas ({\tt pion}) was regarded as insignificant.
(Even if the ionized gas ({\tt pion}) is included within the M4 model was not significant.)

According to the results obtained by the best model (M3 model), an ionized gas exhibited high ionization energy of $\log\xi~({\rm erg~cm~s}^{-1}) = 2.1^{+0.5}_{-0.1}$, high column density of $N_{\rm H} = 3.5^{+1.6}_{-0.9}\times10^{22}~{\rm cm}^{-2}$, and an extremely fast outflow velocity of $v_{\rm out} = -34700^{+400}_{-200}~{\rm km~s}^{-1}$.
Thus, this ionized gas component was identified to be a UFO.
The other ionized gas has a lower ionization parameter of $\log\xi~({\rm erg~cm~s}^{-1}) = 0.51^{+0.19}_{-0.15}$, lower column density of $N_{\rm H} = 8.1^{+0.8}_{-0.6}\times10^{21}~{\rm cm}^{-2}$ and a lower outflow velocity of $v_{\rm out} = -7600 \pm 200~{\rm km~s }^{-1}$.
Thus, this ionized gas component was deemed as a WA.

\begin{figure}
 \centering
 \includegraphics[width=0.9\hsize]{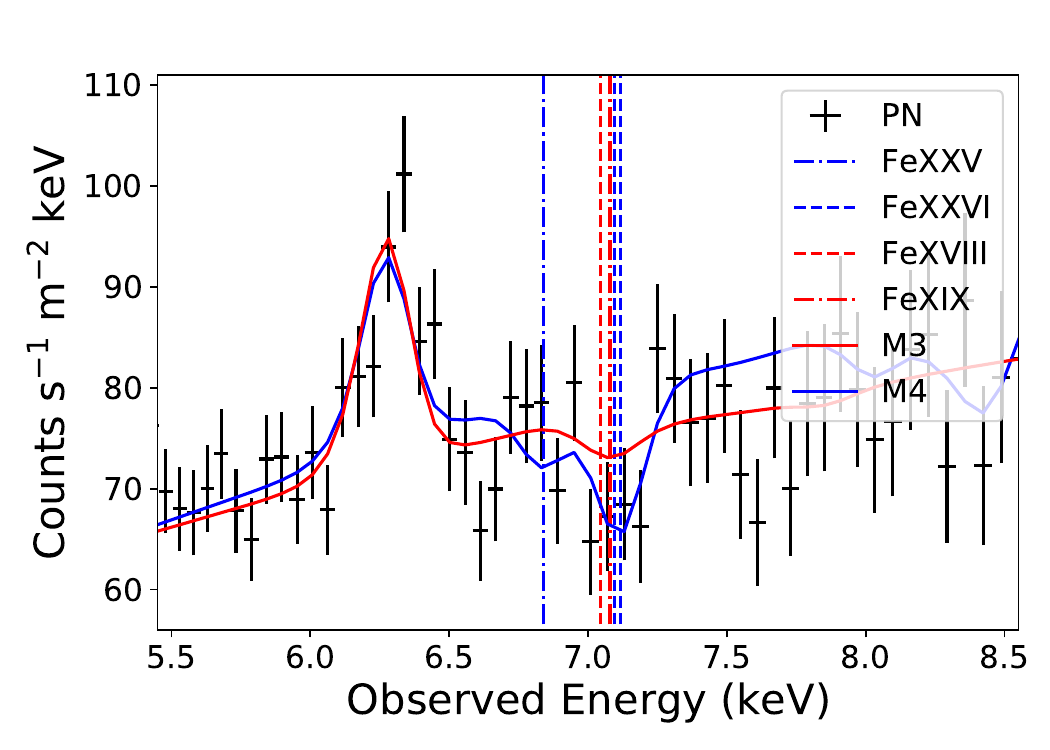}
 \caption{Absorption lines of UFO, wherein the black cross point, red line, blue line, red dashed lines, and blue dashed lines represent the EPIC/pn spectra, M3 model, M4 model, UFO absorption lines ($EW > 2~{\rm eV}$) calculated based on M3 model and UFO absorption lines ($EW >2~{\rm eV}$) calculated based on M3 model respectively.}
 \label{ufoew1}
\end{figure}

\begin{figure}
 \centering
 \includegraphics[width=0.9\hsize]{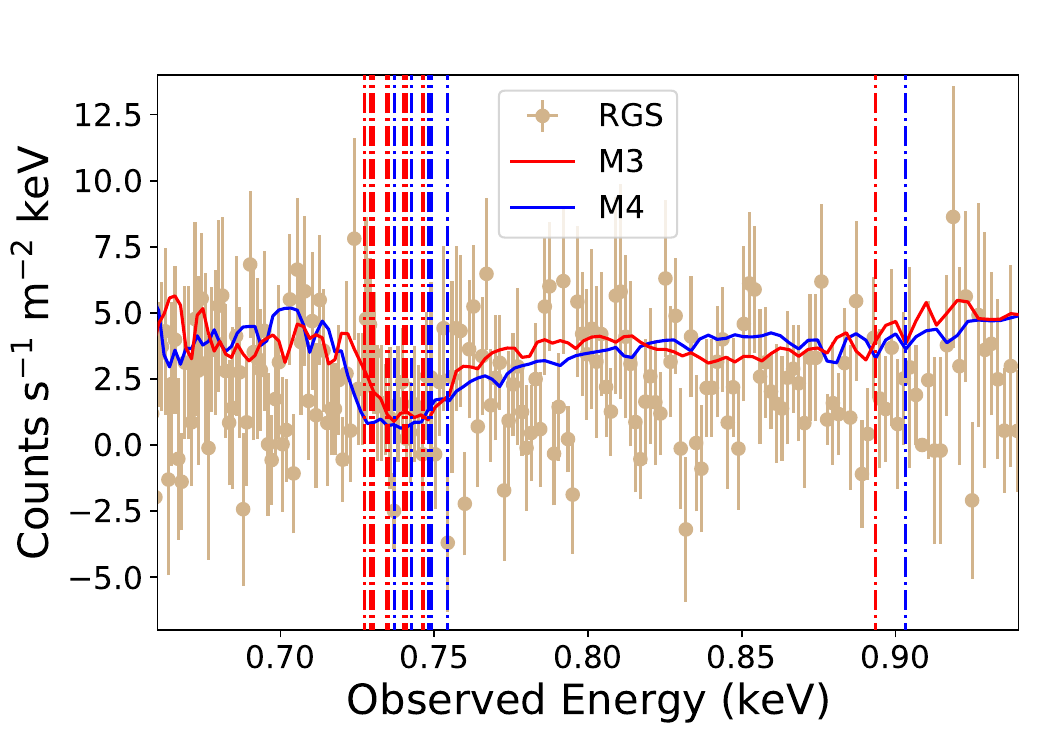}
 \caption{Absorption lines of WA, wherein the orange point, red line, blue line, red dashed lines, and blue dashed lines represent the combined RGS spectra, M3 model, M4 model, WA absorption lines ($EW > 2~{\rm eV}$) calculated based on M3 model and WA absorption lines ($EW > 2~{\rm eV}$) calculated based on M3 model, respectively. RGS spectra is combined by {\tt rgscombine} command. For plotting, the binning factor of the combined RGS spectrum was set to 5.}
 \label{waew1}
\end{figure}

We evaluated the absorption lines of UFOs and WAs.
The UFO absorption lines above 6.0 keV with $EW > 2~{\rm eV}$ and soft X-ray WA absorption lines with $EW > 2~{\rm eV}$are illustrated in Figure \ref{ufoew1} and Figure \ref{waew1}, respectively.
In addition, the information on UFO absorption is likely better identified with presented in Table \ref{ufolisttable}.

\begin{table*}
 \caption{Absorber Search.}
 \label{comt_tau2}
 \centering
 \begin{threeparttable}
\small\begin{tabular}{lcccccccc}
\noalign{\smallskip}\hline\noalign{\smallskip}
 & M1 & M2 & M3 & M4 \\
\noalign{\smallskip}\hline\noalign{\smallskip}
{\tt pow} $\Gamma$ $^{(1)}$ & $1.35\pm0.02$ & $1.48 \pm 0.03$ & $1.57^{+0.03}_{-0.04}$ & $1.38 \pm 0.02$\\
{\tt pow} Norm $^{(2)}$ & $1.77\pm0.05$ & $2.4 \pm 0.1$ & $2.8^{+0.1}_{-0.2}$ & $3.3^{+2.1}_{-1.0}$ \\\noalign{\smallskip}
{\tt refl} Scal $^{(3)}$ & $0.32 \pm 0.05$ & $0.36 \pm 0.06$ & $0.38^{+0.06}_{-0.05}$ & $0.17^{+0.04}_{-0.05}$\\\noalign{\smallskip}
{\tt hot} $N_{\rm H}$ $^{(4)}$ & $0.70^{+0.04}_{-0.03}$ & $0.46^{+0.04}_{-0.03}$ & - & - \\\noalign{\smallskip}
{\tt mbb} Norm $^{(5)}$ & $8^{+4}_{-3}\times10^{9}$ & $8^{+4}_{-3}\times10^{8}$ & $<2\times10^{8}$ & $2^{+3}_{-1}\times10^{6}$ \\\noalign{\smallskip}
{\tt mbb} t $^{(6)}$ & $64\pm2$ & $72\pm3$ & $80^{+20}_{-10}$ & $14 \pm 30$\\\noalign{\smallskip}
{\tt pion} $\log \xi$ $^{(7)}$ & - & $3.1^{+0.1}_{-0.2}$ & $2.1^{+0.5}_{-0.1}$ & $3.1 \pm 0.1$\\\noalign{\smallskip}
{\tt pion} $N_{\rm H}$ $^{(8)}$ & - & $1.7\pm0.3$ & $3.5^{+1.6}_{-0.9}$ & $70^{+60}_{-40}$\\\noalign{\smallskip}
{\tt pion} $v_{\rm out}$ $^{(9)}$ & - & $-21800^{+200}_{-300}$ & $-34700^{+400}_{-200}$ & $-12000^{+200}_{-300}$\\\noalign{\smallskip}
{\tt pion} $f_{\rm cov}$ $^{(10)}$ & - & - & $0.7\pm0.1$ & -\\\noalign{\smallskip}
{\tt pion} $A_{\rm Fe}$ $^{(11)}$ & - & - & - & $4^{+4}_{-1}$\\\noalign{\smallskip}
{\tt pion} $\log \xi$ $^{(7)}$ & - & - & $0.51^{+0.19}_{-0.15}$ & $1.0 \pm 0.1$\\\noalign{\smallskip}
{\tt pion} $N_{\rm H}$ $^{(8)}$ & - & - & $0.81^{+0.08}_{-0.06}$ & $1.1 \pm 0.1$\\\noalign{\smallskip}
{\tt pion} $v_{\rm out}$ $^{(9)}$ & - & - & $-7600 \pm 200$ & $-4300^{+100}_{-200}$\\
\noalign{\smallskip}\hline\noalign{\smallskip}
Total {\it C}-stat / d.o.f. & 2286.8 / 1727 & 2218.8 / 1724 & 2157.1 / 1722 & 2193.7 / 1722 \\
RGS1 {\it C}-stat / bins & 1085.0 / 803 & 1054.9 / 803 & 1020.6 / 803 & 1015.6 / 803 \\
RGS2 {\it C}-stat / bins & 982.7 / 768 & 981.0 / 768 & 969.7 / 768 & 997.8 / 768 \\
EPIC/pn {\it C}-stat/bins & 220.2 / 162 & 183.0 / 162 & 166.8 / 162 & 180.3 / 162 \\
Total Expected {\it C}-stat & $1911.5\pm57.4$ & $1918.0\pm57.5$ & $1917.4\pm57.5$ & $1912.1\pm57.4$ \\
RGS1 Expected {\it C}-stat & 886.1 & 892.8 & 896.2 & 895.5\\
RGS2 Expected {\it C}-stat & 862.6 & 863.2 & 859.0 & 854.5\\
EPIC/pn Expected {\it C}-stat & 162.1 & 162.1 & 162.2 & 162.2\\
$\Delta${\it C}-stat & $-$ & $-68.0$ & $-129.7$ & $-93.1$\\
\noalign{\smallskip}\hline\noalign{\smallskip}
\end{tabular}
\begin{tablenotes}
\item
Fit the total spectra (0.34-10 keV) with a continuum (modified blackbody model $+$ power-law $+$ reflection) and the absorption component.
The cut-off energy of the power law and reflection was set to 120 keV.
$\Delta${\it C}-stat: Compare the {\it C}-stat with M1 model.
M1: With {\tt hot} model.
M2: With one {\tt pion} model and one {\tt hot} model.
M3, M4: With two {\tt pion} models.
(1) Photon index of the X-ray power-law component.
(2) Normalization of the X-ray power-law component in $10^{51} ~{\rm photons~s^{-1}~keV^{-1}}$ at 1 keV.
(3) Scale parameter of the reflection component.
(4) Column density $N{\rm _H}$ of neutral ISM gas component in the host galaxy of the AGN in $10^{22}$ cm$^{-2}$.
(5) Normalization of the modified blackbody in $10^{34} ~{\rm cm^{0.5}}$.
(6) Temperature of the modified blackbody in eV.
(7) Logarithm of the ionization parameter $\xi$ of the ionized wind components in ${\rm erg ~cm ~s^{-1}}$. 
(8) Column density $N{\rm _H}$ of the ionized wind components in $10^{22}$ cm$^{-2}$. 
(9) Velocity of the outflowing ionized wind components in ${\rm km ~s^{-1}}$.
(10) Covering factor of the {\tt pion} component.
(11) Fe abundance of the pion component.
\end{tablenotes}
\end{threeparttable}

\end{table*}

\begin{figure}
 \centering
 \includegraphics[width=0.9\hsize]{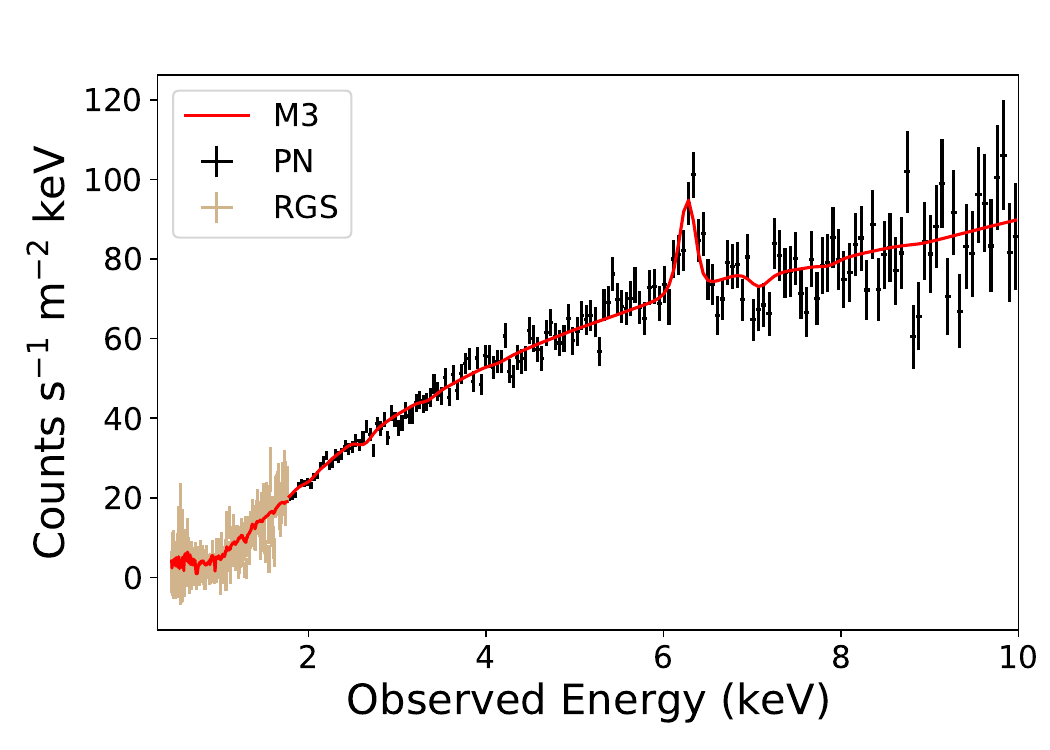}
 \caption{All spectra (RGS + EPIC/pn) and model curve, wherein the black cross point, orange point, and red line represent EPIC/pn spectra, combined RGS spectra, and the M3 model, respectively. For the EPIC/pn spectra, we used the {\tt obin} command for optimal binning. RGS spectra were combined by {\tt rgscombine} command and the binning factor was set to 5.}
 \label{absall}
\end{figure}

\begin{figure}
 \centering
 \includegraphics[width=0.9\hsize]{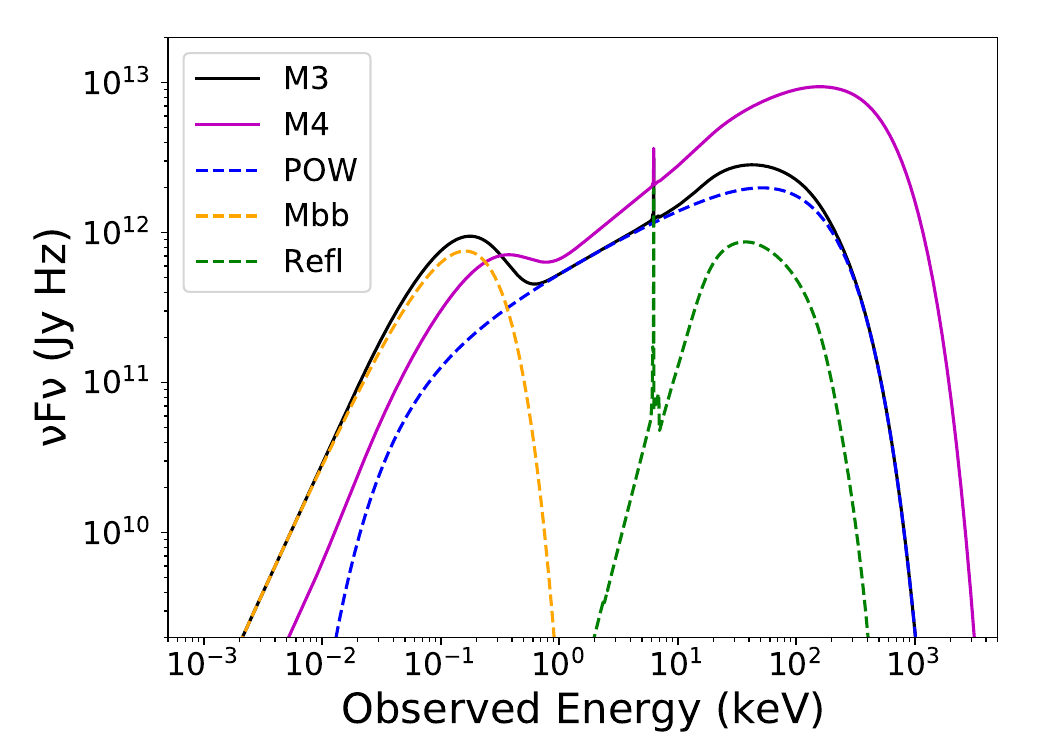}
 \caption{SED curve of the model, wherein the black line, blue line, orange line, green line, and magenta line represent the M3 model SED, power-law model components, modified blackbody model, reflection model components, and the M4 model, respectively.}
 \label{2pionsed}
\end{figure}

\begin{table*}
 \caption{The UFO absorption lines from the M3 model and the M4 model ($EW > 2~{\rm eV}$).}
 \label{ufolisttable}
 \centering
 \begin{threeparttable}
  \begin{tabular}{lccc}
\hline\noalign{\smallskip}
Absorption line & Rest-frame Energy (keV) & $EW$ (eV) & Observed Energy (keV)\\
\noalign{\smallskip}\hline\noalign{\smallskip}
\multicolumn{4}{c}{M3 model ($v_{\rm out} = -34700~{\rm km/s}$)}\\ \noalign{\smallskip}
Fe {\sc xix} & 6.4656 & 3.5 & 7.0794\\
Fe {\sc xviii} & 6.4347 & 3.4 & 7.0456\\ \noalign{\smallskip}
\multicolumn{4}{c}{M4 model ($v_{\rm out} = -12000~{\rm km/s}$)}\\ \noalign{\smallskip}
Fe {\sc xxvi} & 6.9732 & 23.4 & 7.1180\\
Fe {\sc xxvi} & 6.9517 & 20.4 & 7.0960\\
Fe {\sc xxv} & 6.7004 & 18.2 & 6.8395\\
\noalign{\smallskip}\hline\noalign{\smallskip}
\end{tabular}
\begin{tablenotes}
\item
These lines are based on the M3 model ($v_{\rm out} = -34700~{\rm km/s}$) and M4 model ($v_{\rm out} = -12000~{\rm km/s}$) for each. 
The red dashed line and blue dashed-dotted line in Fig. \ref{ufoew1} are the absorption lines shown here. 
\end{tablenotes}
 \end{threeparttable}

\end{table*}

\section{Discussion}
\subsection{UFO}
As reported in prior research, 3C 390.3 and 4C $+$74.26 are the only two galaxies where both UFOs and WAs have been observed, with Mrk 6 being the third case considered in this study (Table \ref{UFOcompare}).
Compared to the Mrk 6 UFOs with 3C 390.3 and 4C $+$74.26, the ionization was lower, and the column density and outflow velocity are comparable (Table \ref{UFOcompare}). 
Although, both 3C 390.3 and 4C $+$74.26 have lower limits for the column density \citep{Tombesi2014}.

\begin{table}
 \caption{Comparison of parameter of UFOs in 3C 390.3 and 4C $+$74.26.}
 \label{UFOcompare}
 \centering
 \begin{threeparttable}
\small\begin{tabular}{lcccccccc}
\noalign{\smallskip}\hline\noalign{\smallskip}
 & Mrk 6 & 3C 390.3$^*$ & 4C $+$74.26$^*$ \\
\noalign{\smallskip}\hline\noalign{\smallskip}
$\log \xi$ $^{(1)}$ & $2.1^{+0.5}_{-0.1}$ & $5.6^{+0.2}_{-0.8}$ & $4.62\pm0.25$ \\
$N_{\rm H}$ $^{(2)}$ & $3.5^{+1.6}_{-0.9}$ & $>3$ & $>4$ \\
$v_{\rm out}$ $^{(3)}$ & $-34700^{+400}_{-200}$ & $-44000\pm1000$ & $-13000\pm2000$ \\
\noalign{\smallskip}\hline\noalign{\smallskip}
\end{tabular}
\begin{tablenotes}
\item
*: \citet{Tombesi2014}. The {\tt xabs} model was used for fitting.
(1) Logarithm of the ionization parameter $\xi$ of the ionized wind components in ${\rm erg ~cm ~s^{-1}}$. 
(2) Column density $N{\rm _H}$ of the ionized wind components in $\times 10^{22}~$cm$^{-2}$. 
(3) Velocity of the outflowing ionized wind components $\rm km ~s^{-1}$.

\end{tablenotes}
\end{threeparttable}
\end{table}

The ionization parameter is defined as $\xi = L_{\rm ion} / n_{\rm H}r^2$ \citep{Tarter1969} using the hydrogen density $n_{\rm H}$, disk wind distance $r$, and luminosity of the light source $L_{\rm ion}$. 
Thus, the distance of the disk wind was calculated as 
\begin{eqnarray}\label{eqr1}
r=\left({L_{\rm ion} \over n_{\rm H}\xi }\right)^{0.5}
\end{eqnarray}
Based on the M3 model result, the UFO ionization parameter was $\log_{10}\xi = 2.1^{+0.5}_{-0.1}~{\rm erg~cm~s}^{-1}$ and the non-filtered (before getting absorbed by the UFO {\tt pion} component) ionized luminosity was calculated to $L_{\rm ion} / \xi = 3.2\times10^{41}~{\rm cm^{-1}}$.
Therefore, the distance of the UFO from the SMBH could be estimated as
\begin{eqnarray}
r&=&\left({3.2\times10^{41}\over10^{2.1}\times n_{\rm H}}\right)^{0.5}\nonumber\\
&=& 5.0\times10^{15}\left(\frac{n_{\rm H}}{10^{8} {\rm cm}^{-3}}\right)^{-0.5}~{\rm cm} 
\end{eqnarray}
This is equivalent to $95\left(\frac{n_{\rm H}}{10^{8} {\rm cm}^{-3}}\right)^{-0.5}~R_{\rm S}$, where the Schwarzschild radius $R_{\rm S} = 2GM_{\rm BH} / c^2 = 5.3\times10^{13}~{\rm cm}$.
Assuming that the wind has a higher velocity than the escape velocity.
Therefore, the lower limit of the distance of the disk wind was evaluated as 
\begin{eqnarray}\label{eqr2}
r_{\rm min}={2GM_{\rm BH} \over v_{\rm out}^2}
\end{eqnarray}
using the gravitational constant $G$, SMBH mass $M_{\rm BH}$, and outflow velocity $v_{\rm out}$.
Based on the analysis results, the outflow velocity of the UFO was $v_{\rm out} = 34700~{\rm km~s^{-1}}$, and therefore, the lower limit of the UFO distance was calculated as $r_{\rm min} = 4.0\times10^{15}~{\rm cm}~( = 75~R_{\rm S})$.
Assuming that the disk wind width $\Delta r$ did not exceed the distance $r$ to the SMBH, the upper limit of the disk wind distance was computed as 
\begin{eqnarray}\label{eqr3}
r_{\rm max}={{L_{\rm ion} C_{\rm v} \over \xi N_{\rm H} }}
\end{eqnarray}
considering the covering factor $C_{\rm v}$ of the disk wind \citep{Wang2022}.
M3 model assumed a covering factor of $C_{\rm v}=1$, such that the upper limit of the UFO distance was $r_{\rm max} = 7.2\times10^{16}~{\rm cm}~( = 1.3\times10^{3}~R_{\rm S})$, considering a column density of $N_{\rm H} = 3.5\times10^{22}~{\rm cm}^{-2}$.
Therefore, the UFOs of Mrk 6 originated at $95\left(\frac{n_{\rm H}}{10^{8} {\rm cm}^{-3}}\right)^{-0.5}~R_{\rm S}$, and the range of the UFOs is $7.5\times10^1-1.3\times10^{3}~R_{\rm S}$.
Based on observational studies, UFOs are considered to originate in the $10^2-10^4~R_{\rm S}$ \citep[e.g.][]{Gofford2015}, and according to simulation research, they originate in the $\sim30~R_{\rm S}$ \citep[e.g.][]{Nomura2016}.
Given the observational and model uncertainties, here you can say that the range is consistent with what expected from disk winds.

Furthermore, we evaluated the range of hydrogen density from that of UFOs. 
Based on the definition of ionization, the density of the disk wind can be evaluated as 
\begin{eqnarray}\label{eqr4}
n_{\rm H} = {L_{\rm ion} \over \xi r^2}
\end{eqnarray}
The range of the UFOs was $4.0\times10^{15}-7.2\times10^{16}~{\rm cm}$, and thus, the range of the hydrogen density was calculated as $4.8\times10^{5}-1.6\times10^{8}~{\rm cm}^{-3}$.

\subsection{WA}
WAs are often reported to contain multiple components, e.g., NGC 3516 has 3 WAs \citep{Mehdipour2010}, NGC 7469 has 4 WAs \citep{Mehdipour2018}, NGC 5548 has 6 WAs \citep{Mao2017}, NGC 3783 has 9 WAs \citep{Mao2019} and NGC 3227 has 4 WAs \citep{Wang2022}.
Both 3C 390.3 and 4C $+$74.26 have two WAs components each \citep{Mehdipour19}.
However, for Mrk 6, the second WA component is not significant.
Thus, \citet{Mehdipour19} used only one WA component along with a neutral absorption component.
Compared to the Mrk 6 WAs with 3C 390.3 and 4C $+$74.26, the ionization was low and the outflow velocity was high, however, the column densities did not vary significantly (Table \ref{WAcompare}).

\begin{table*}
 \caption{Comparison of parameters of WAs in the 3C 390.3 and 4C $+$74.26.}
 \label{WAcompare}
 \centering
 \begin{threeparttable}
\small\begin{tabular}{lcccccccc}
\noalign{\smallskip}\hline\noalign{\smallskip}
 & Mrk 6 & 3C 390.3$^*$ & 4C $+$74.26$^*$ \\
\noalign{\smallskip}\hline\noalign{\smallskip}
$\log \xi$ $^{(1)}$ & $0.51^{+0.19}_{-0.15}$ & $1.63\pm0.11$, $2.77\pm0.06$ & $1.69\pm0.04$, $2.46\pm0.04$ \\
$N_{\rm H}$ $^{(2)}$ & $8.1^{+0.8}_{-0.6}$ & $0.37\pm0.06$, $1.1\pm0.5$ & $3.6\pm0.6$, $6.8\pm0.8$ \\
$v_{\rm out}$ $^{(3)}$ & $-7600 \pm 200$ & $-1500\pm60$, $50\pm100$ & $-1490\pm90$, $-3000\pm500$ \\
\noalign{\smallskip}\hline\noalign{\smallskip}
\end{tabular}
\begin{tablenotes}
\item
*: \citet{Mehdipour19}. The {\tt pion} model was used for fitting.
(1) Logarithm of the ionization parameter $\xi$ of the ionized wind components in ${\rm erg ~cm ~s^{-1}}$. 
(2) Column density $N{\rm _H}$ of the ionized wind components in $\times 10^{22}~$cm$^{-2}$. 
(3) Velocity of the outflowing ionized wind components $\rm km ~s^{-1}$.
\end{tablenotes}
\end{threeparttable}
\end{table*}

Based on the results of the analysis with the M3 model, the ionization parameter of WA was $\log_{10} \xi~({\rm erg~cm~s}^{-1}) = 0.51^{+0.19}_{-0.15}$ and the filtered (after getting absorbed by the UFO {\tt pion} component and before getting absorbed by the WA {\tt pion} component) ionized luminosity was calculated to $L_{\rm ion} / \xi = 1.0\times10^{43}~{\rm cm^{-1}}$.
Therefore, with Equation (\ref{eqr1}), the distance of WA from SMBH was 
\begin{eqnarray}
r&=& \left({1.0\times10^{43}\over10^{0.5}\times n_{\rm H}}\right)^{0.5}\nonumber\\
&=& 5.6\times10^{18}\left(\frac{n_{\rm H}}{10^{5} {\rm cm}^{-3}}\right)^{-0.5}~{\rm cm}
\end{eqnarray}
which is equivalent to $1.8\left(\frac{n_{\rm H}}{10^{5} {\rm cm}^{-3}}\right)^{-0.5}~{\rm pc}$.
The fitting results exhibited that the outflow velocity of the WA was $v_{\rm out} = 7600~{\rm km~s^{-1}}$, and the lower limit of the WA distance was $r_{\rm min} = 8.3\times10^{16}~{\rm cm}~(=2.7\times10^{-2}~{\rm pc}$) using Equation (\ref{eqr2}).
As the column density of the WA was $N_{\rm H}=8.1^{+0.8}_{-0.6}\times10^{21}~{\rm cm}^{-3}$ according to the analysis and the coverage factor was assumed to be $C_{\rm v}=1$, the upper limit of the WA distance was calculated to be $r_{\rm max} = 3.9\times10^{20}~{\rm cm}~(=1.3\times10^{2}~{\rm pc})$ using Equation (\ref{eqr3}).
Therefore, the WAs of Mrk 6 originated at $1.8\left(\frac{n_{\rm H}}{10^{5} {\rm cm}^{-3}}\right)^{-0.5}~{\rm pc}$, and the range of the WAs was $2.7\times10^{-2}-1.3\times10^{2}~{\rm pc}$.
WAs are considered to originate in the $0.1-1000$ pc \citep{Laha2021}, and the present findings are consistent with this result.

Similar to the case with UFOs, the hydrogen density of the WA was calculated as well.
The WA distance range was calculated to be $8.3\times10^{16}-3.9\times10^{20}~{\rm cm}$, such that the hydrogen density of WA was calculated as $2.1\times10^1-4.6\times10^{8}~{\rm cm }^{-3}$ using Equation (\ref{eqr4}).
The hydrogen number density of WAs was estimated to be $\sim 10^{4}-10^{11}~{\rm cm}^{-3}$ \citep[][and references therein]{Laha2021}, and the present result was wider than the range.

\subsection{XRISM simulation}
As the EPIC/pn instrument of {\it XMM-Newton} exhibited a low energy resolution, the absorption lines could not be clearly observed (Figure \ref{absall}) and the ionized wind parameters (ionization parameter, hydrogen column density, and outflow velocity) yielded large errors (Table \ref{comt_tau2}).
However, XRISM/Resolve \citep{Tashiro2020, XRISM2022} provides an ultrahigh energy resolution, which facilitates the prominent observation of the absorption line and reduces the uncertainties of the parameters and XRISM to verify whether Mrk 6 contained only one WA component.
Thereafter, we simulated the observation in XRISM/Resolve based on the M3 model, which was determined to be the best-fit model.
The XRISM/Resolve response file was used with an energy resolution of 7 eV version 2019a\footnote{\url{https://xrism.isas.jaxa.jp/research/proposer/obsplan/response/index.html}}.
For the simulations and fitting, the SPEX code was used, and the exposure time was set to 150 ks.
In the analysis, the energy range was set to $0.34-10.0$ keV.
For comparison, we simulated the M4 model to obtain results with Fe {\sc xxv}/{\sc xxvi} absorption lines at $\sim7$ keV.

\begin{figure}
 \centering
 \includegraphics[width=0.9\hsize]{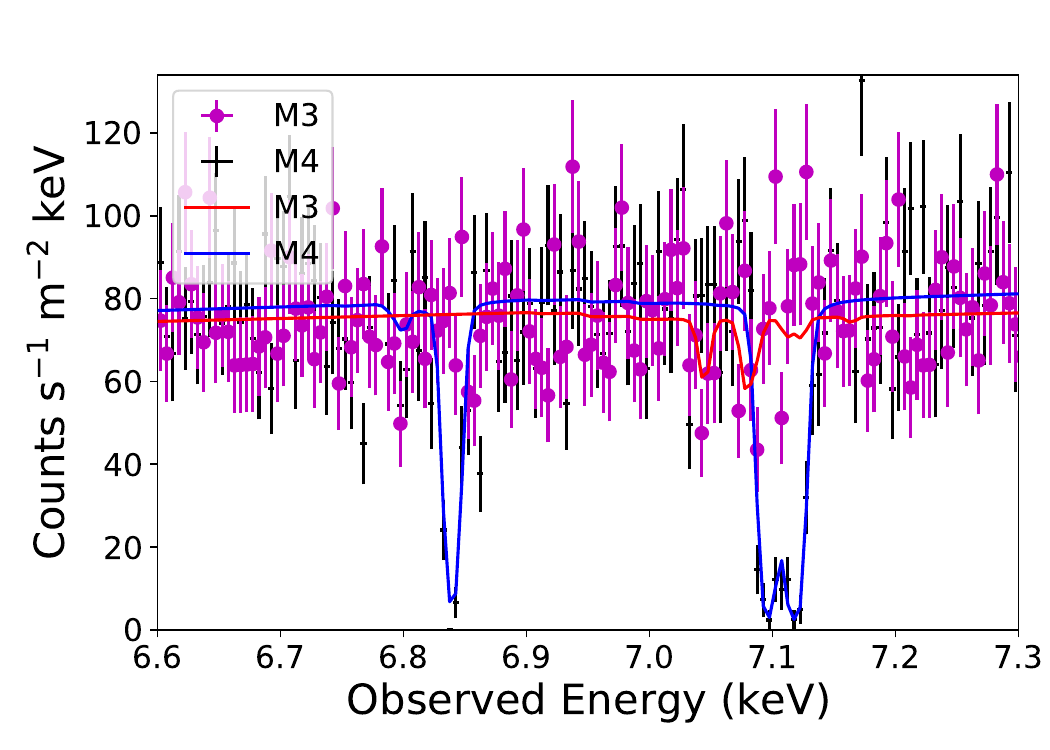}
 \caption{The $6.6-7.3$ keV energy band simulated spectra and model curve. The magenta point, black cross point, blue line, and red dashed line represent the XRISM simulated spectra based on the M3 model, and the XRISM simulated spectra based on the M4 model, the M3 model, and the M4 model, respectively. For plotting, the binning factor was set to 10 for the simulated XRISM spectra.}
 \label{xrismfe}
\end{figure}

\begin{figure}
 \centering
 \includegraphics[width=0.9\hsize]{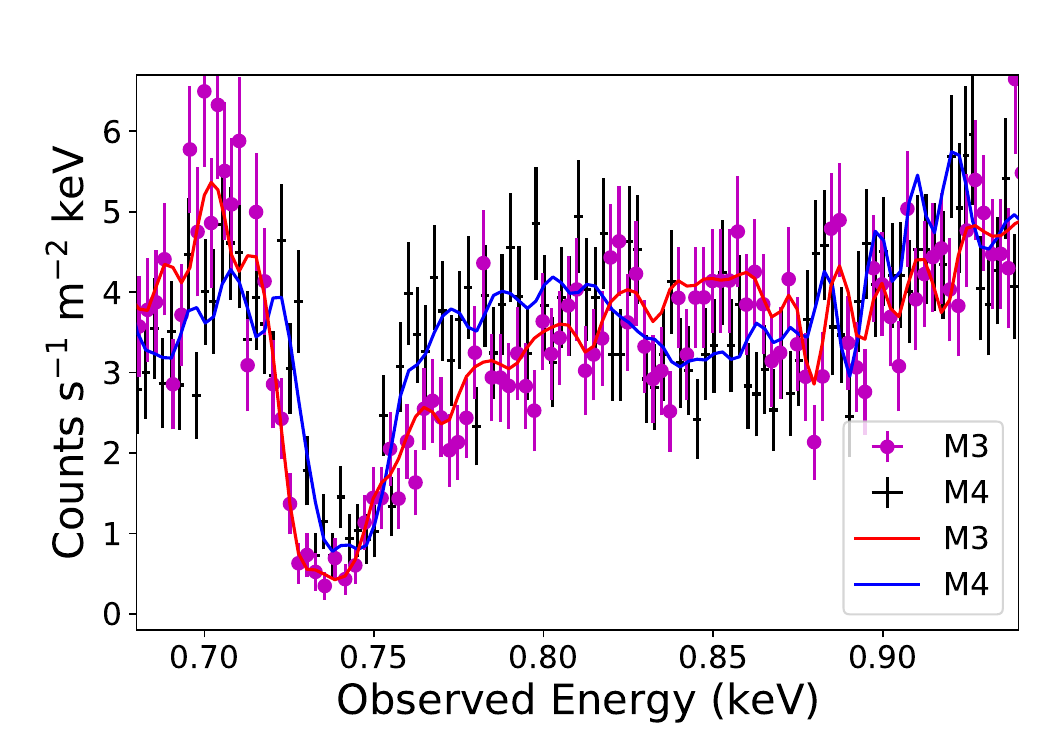}
 \caption{The $0.68-0.94$ keV energy band simulated spectra and model curve. The magenta point, black cross point, blue line, and red dashed line represent the XRISM simulated spectra based on the M3 model, and the XRISM simulated spectra based on the M4 model, the M3 model, and the M4 model, respectively. For plotting the simulated XRISM spectra, we used the {\tt obin} command for optimal binning}
 \label{xrismsoft}
\end{figure}

Based on the findings listed in Table \ref{FeKworse}, the errors of the disk wind parameters in both the M3 and M4 models were smaller than those reported in Table \ref{comt_tau2}.
In particular, the error in the column density $N_{\rm H}$, which is a vital parameter for the significant detection of the disk wind, was reduced.
The $\sim 7$ keV absorption lines can be more prominently observed (Figure \ref{xrismfe}).
In addition, the soft X-ray spectra error bar was more accurate (Figure \ref{xrismsoft}).
Therefore, XRISM/Resolve can more accurately detect the disk wind.
As observed in Figure \ref{xrismfe}, the Fe {\sc xix}/{\sc xviii} absorption lines are weak, and the Fe {\sc xxv}/{\sc xxvi} absorption lines are strong.
Furthermore, the Fe {\sc xxv}/{\sc xxvi} and Fe {\sc xix}/{\sc xviii} absorption lines appeared prominently at various energies. 
Thus, we can employ the XRISM/Resolve to decide whether the $\sim7$ keV absorption lines in Mrk 6 can be attributed to Fe {\sc xxv}/{\sc xxvi} or Fe {\sc xix}/{\sc xviii}.

\begin{table}
 \caption{XRISM/Resolve simulation.}
 \label{FeKworse}
 \centering
 \begin{threeparttable}
\small\begin{tabular}{lcccccccc}
\noalign{\smallskip}\hline\noalign{\smallskip}
  & M3 & M4 \\
\noalign{\smallskip}\hline\noalign{\smallskip}
{\tt pow} $\Gamma$ $^{(1)}$ & $1.56\pm0.02$ & $1.40\pm0.01$\\
{\tt pow} Norm $^{(2)}$ & $2.8\pm0.1$& $3.6^{+0.5}_{-0.4}$\\
{\tt refl} Scal $^{(3)}$ & $0.37\pm0.02$ & $0.16\pm0.02$\\
{\tt mbb} Norm $^{(4)}$ & $1.1^{+0.6}_{-0.4}\times10^{8}$ & $3.1^{+1.0}_{-0.7}\times10^{6}$\\
{\tt mbb} t $^{(5)}$ & $82^{+8}_{-6}$ & $130\pm10$\\
{\tt pion} $\log \xi$ $^{(6)}$ & $2.09^{+0.04}_{-0.03}$ & $3.10\pm0.04$\\\noalign{\smallskip}
{\tt pion} $N_{\rm H}$ $^{(7)}$ & $3.6\pm0.3$ & $80^{+20}_{-10}$\\\noalign{\smallskip}
{\tt pion} $v_{\rm out}$ $^{(8)}$ & $-34650\pm40$ & $-12040\pm10$\\
{\tt pion} $f_{\rm cov}$ $^{(9)}$ & $0.67\pm0.02$ & -\\
{\tt pion} $A_{\rm Fe}$ $^{(10)}$ & - & $3.3^{+0.5}_{-0.4}$\\
{\tt pion} $\log \xi$ $^{(6)}$ & $0.46^{+0.06}_{-0.05}$ & $0.97\pm0.03$\\
{\tt pion} $N_{\rm H}$ $^{(7)}$ & $0.80\pm0.02$ & $1.13^{+0.03}_{-0.02}$\\
{\tt pion} $v_{\rm out}$ $^{(8)}$ & $-7800\pm100$ & $-4400\pm100$\\
\noalign{\smallskip}\hline\noalign{\smallskip}
Total {\it C}-stat / d.o.f. & 4095.6 / 3935 & 3920.9 / 3935\\
Total Expected {\it C}-stat & $3992.1\pm90.1$ & $3995.7\pm90.1$ \\
\noalign{\smallskip}\hline\noalign{\smallskip}
\end{tabular}
\begin{tablenotes}
\item
Fit the total spectra (0.34-10 keV) with a continuum (modified blackbody model $+$ power-law $+$ reflection) and the absorption component.
The cut-off energy of the power law and reflection was set to 120 keV.
(1) Photon index of the X-ray power-law component.
(2) Normalization of the X-ray power-law component in $10^{51} ~{\rm photons~s^{-1}~keV^{-1}}$ at 1 keV.
(3) Scale parameter of the reflection component.
(4) Normalization of the modified blackbody in $10^{34} ~{\rm cm^{0.5}}$.
(5) Temperature of the modified blackbody in eV.
(6) Logarithm of the ionization parameter $\xi$ of the ionized wind components in ${\rm erg ~cm ~s^{-1}}$. 
(7) Column density $N{\rm _H}$ of the ionized wind components in $10^{22}$ cm$^{-2}$. 
(8) Velocity of the outflowing ionized wind components) in ${\rm km ~s^{-1}}$.
(9) Covering factor of the {\tt pion} component.
(10) Fe abundance of the pion component.

\end{tablenotes}
\end{threeparttable}

\end{table}

\section{Summary}
This study analyzed the disk wind of the radio galaxy Mrk 6 using {\it XMM-Newton} archival data with the SPEX code. 
Based on the results of the M3 model, the existence of both UFOs and WAs was suggested in Mrk 6.
In addition to the radio galaxies 3C 390.3, 4C +74.26, Mrk 6 is the third radio galaxy in which both UFOs and WAs have been observed.
Compared to the Mrk 6 disk wind with 3C 390.3 and 4C $+$74.26, the UFOs exhibited low ionization, however, the UFO and WA parameters are broadly consistent with the ones reported for the other two radio galaxies.
The distance from SMBH of UFOs and WAs are $95\left(\frac{n}{10^{8} {\rm cm}^{-3}}\right)^{-0.5}~R_{\rm S}$ and $1.8\left(\frac{n}{10^{5} {\rm cm}^{-3}}\right)^{-0.5}~{\rm pc}$, respectively.
These distances were consistent with the typical ones. 
The M3 model attributes the ~7 keV absorption lines to Fe {\sc xix}/{\sc xviii}.
For comparison, the M3 model was re-fitted with varying parameters to obtain a model that attributed the ~7 keV absorption lines to Fe {\sc xxv}/{\sc xxvi} (M4 model) and it may represent a local minimum of the C-stat.
Although the $\sim7$ keV absorption lines can be attributed to Fe {\sc xxv}/{\sc xxvi} absorption lines, the spectra derived herein suggested that the $\sim7$ keV absorption lines are actually Fe {\sc xix}/{\sc xviii} absorption lines.
Finally, we performed simulations of observations with the XRISM satellite and determined that the errors in the {\tt pion} model parameters were extremely small using XRISM. 
Therefore, XRISM can be used to detect significant disk winds. 
In addition, based on XRISM simulations we investigated whether the $\sim7$ keV absorption lines in Mrk 6 were caused by Fe {\sc xxv}/{\sc xxvi} or Fe {\sc xix}/{\sc xviii}.

\section*{Acknowledgements}
The authors thank the anonymous referee for their useful comments to improve the quality of this work.
This work is based on observations obtained with {\it XMM-Newton}.
This work was supported by JST, the establishment of university fellowships toward the creation of science technology innovation, Grant Number JPMJFS2129.
\section*{Data Availability}
A reproduction package is available at Zenodo DOI: \href{https://zenodo.org/doi/10.5281/zenodo.10394007}{10.5281/zenodo.10394008}. This package includes data and scripts used to reproduce the fitting results and figures presented here.



\bibliographystyle{mnras}








\bsp	
\label{lastpage}
\end{document}